\documentclass[conference]{IEEEtran}

\newtheorem{lemma}{Lemma}

\newcommand{\qudit}[1]{\left\vert #1 \right\rangle}
\newcommand{\rqudit}[1]{\left\langle #1 \right\vert}

\newcommand{\Z}{\mathbb{Z}}

\newcommand{\C}{\mathbb{C}}

\usepackage{amssymb}\usepackage[official]{eurosym}

\usepackage{amssymb, color}

\begin{document}
\title{On interchanging the states of a pair of qudits}
\author{\authorblockN{Colin Wilmott$^1$ and Peter Wild$^2$}
\authorblockA{
$^1$School of Mathematical Sciences, University College Dublin,
Dublin 4, Ireland\\ \emph{Electronic Address:}
cmwilmott@maths.ucd.ie\\ $^2$ Department of Mathematics, Royal
Holloway, Egham, Surrey, TW20 0EX, UK\\\emph{Electronic Address:}
P.Wild@rhul.ac.uk }} \maketitle

\begin{abstract}
The qubit {\small{SWAP}} gate has been shown to be an integral
component of quantum circuitry design.  It permutes the states of
two qubits and allows for the storage quantum information,
teleportation of atomic or ionic states,   and is a fundamental
element in the circuit implementation of Shor's algorithm. We
consider the problem of generalising the {\small{SWAP}} gate
beyond the qubit setting. We show that quantum circuit
architectures completely described by instances of the
\small{CNOT} gate can not implement a transposition of a pair of
qudits for dimensions $d \equiv 3 \ (\textrm{mod} \ 4)$. This is
of interest to the question of construction a generalised quantum
\small{SWAP} gate. The task of constructing generalised SWAP gates
based on transpositions of qudit states is argued in terms of the
signature of a permutation.
\end{abstract}

\section{Introduction}
The crux of successful quantum computation is the
implementation of multiple quantum  gates. The most elementary of
multiple quantum gates is to consider some unitary operator
 $U$ within a controlled-$U$ two qubit operation. The corresponding  transformation  given
 by  transformation is written as $\qudit{0}\rqudit{0} \otimes I +\qudit{1}\rqudit{1}\otimes U$
 where the $I$  operation represents the identity transformation. This controlled two qubit
  operator is so called since the application of $U$ on the second qubit  is decided by the state
  of the first qubit. The classic controlled-$U$ gate  is the controlled-{\small{NOT}} ({\small{CNOT}})
  gate and its action with respect to the computational basis is given
as $\qudit{x}\qudit{y} \mapsto \qudit{x}\qudit{y\oplus x}$ where
$\oplus$ represents addition modulo 2. The {\small{CNOT}} gate
plays an important role in quantum computation (DiVincenzo
(1998)). It is the quantum mechanical analogue of the classical
connective {\small{XOR}} gate and is a principle component for
universal computations. It can be used to produce maximally
entangled states similar to the set of EPR pairs (Nielsen and
Chuang (2000)). Furthermore, the controlled-{\small{NOT}} gate
acts as a {measurement} gate (Deutsch (1989)) and   provides a
basis for a so-called {nondemolition} measurement (Chuang and
Yamamoto (1996)) that permits the construction of a syndrome table
as used in error detection and correction.

The quantum network approach to computation resembles the
classical procedure to computing (Vlasov (2003)) where quantum
circuits are formed from a composition of quantum states, quantum
gates and quantum wires (Nielsen and Chuang (2000)). Computations
are described within the Hilbert space ${\cal{H}} =
{(\C^{2})}^{\otimes n}$ of $n$ qubits where each horizontal
quantum circuit wire corresponds to the individual $\C^{2}$
subspaces. Vertical wires in a quantum circuit represent the
\emph{coupling} of arbitrary pairs of quantum gates in a manner
similar to a controlled-$U$ gate. The \emph{depth} of a circuit
refers to the maximum number of gates required to effect necessary
state changes. The \emph{width} of a circuit is the maximum number
of gates in operation in any one time frame. Quantum computations
are then a finite sequence of quantum gates set along the quantum
wires to effect suitable transformations. Unfortunately, there are
only a handful of quantum gates that can be experimentally
realised within the coherence time of their systems (Vatan and
Williams (2004)). Those gates that have been experimentally
demonstrated  are said to be elements of the quantum gate library.
Barenco \emph{et al.} (1995) showed that any quantum operation on
a set of $n$-qubits can be restricted to a composition of
{\small{CNOT}}, and single qubit gates. For this reason, we say
that the  qubit gate library consisting of single qubit gates and
{\small{CNOT}} is universal. Furthermore, it has become standard
in quantum information to express any $n$-qubit quantum operation
as a composition of single qubit gates and {\small{CNOT}} gates.
Consequently, the {\small{CNOT}} gate has acquired special status
as the hallmark of multiqubit control (Vidal and Dawson (2004)).

\begin{figure}
\setlength{\unitlength}{0.08cm} \hspace*{95mm} \hskip-4.5em
\begin{picture}(40,40)(63,0)
\put(0,20){\line(1,0){40}} \put(0,30){\line(1,0){40}}
\put(10,18){\line(0,1){11}} \put(10,20){\circle{4}}
\put(10,30){\circle*{2}} \put(20,20){\line(0,1){12}}
\put(20,20){\circle*{2}} \put(20,30){\circle{4}}
\put(30,18){\line(0,1){11}} \put(30,20){\circle{4}}
\put(30,30){\circle*{2}}
\put(-7,29){$\qudit{\psi}$} \put(-7,19){$\qudit{\phi}$}
\put(44,29){$\qudit{\phi}$} \put(44,19){$\qudit{\psi}$}
\end{picture}
\vskip-3em\caption{Quantum circuit swapping two
qubits.}\label{swap}
\end{figure}
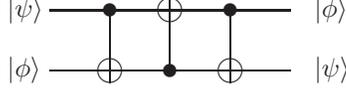

Researchers in universal circuit  constructions have done
considerable work optimising their constructions (Nielsen (2005)).
In particular,  Vatan and Williams (2004) construct a quantum
circuit for a general two-qubit operation that requires at most
three {\small{CNOT}} gates and fifteen one-qubit gates and show
that their construction is   optimal. Crucial to this result is
the demand that the quantum circuit for the two-qubit
{\small{SWAP}} gate requires at least three {\small{CNOT}} gates.
Fig.~\ref{swap} illustrates a quantum circuit swapping the states
of two qubits; system ${\mathcal{A}}$ begins in the state
$\qudit{\psi}$ and ends in the state $\qudit{\phi}$ while system
${\mathcal{B}}$ begins in the state $\qudit{\phi}$ and ends in the
state $\qudit{\psi}$. The {\small{SWAP}} gate has become an
integral feature of the circuitry design of the quantum Fourier
transform where it  can be used to store quantum information, to
teleport atomic or ionic states (Liang  and  Li (2005)). It is
also a fundamental element in the circuit implementation of Shor's
algorithm (Fowler \emph{et al.} (2004)). More recently, a scheme
to realise the quantum {\small{SWAP}} gate between flying and
stationary qubits has been presented by Liang and Li (2005) where
maintained that experimentally realising the quantum
{\small{SWAP}} gate is a necessary condition for the
networkability of quantum computation.

Most often it is assumed that a quantum computer is predicated on
a collection of  qubits. However, there has been the view to
generalise to $d$-level, or \emph{qudit}, quantum mechanical
systems. In the context of information processing, it may be
argued  that there are advantages in moving from the qubit
paradigm to the qudit paradigm. For instance, as the entropy of a
message depends on the alphabet used it ought to be that
increasing the alphabet size should allow for the construction of
better error-correcting codes (Grassl \emph{et al.} (2003)). It
has also been pointed out that a quantum system composed of a pair
of three dimensional subsystems shows new features when compared
to a two-qubit system (Grassl \emph{et al.} (2003)).

We seek to establish conditions for generalising the  quantum
{\small{SWAP}} gate  resulting through instances of the
{\small{CNOT}} gate. We give the following results.

\section{Preliminaries}\label{pre}
Consider the set $\textrm{N} = \{1,2,\dots,n\}$ and let $\sigma:
\textrm{N} \mapsto \textrm{N}$ be a bijection. We say $\sigma =
\left[%
\begin{array}{cccc}
 1&2&\dots&n \\
  i_1&i_2&\dots&i_n \\
\end{array}%
\right]$
, where $i_k \in \textrm{N}$
is the image of $k \in \textrm{N}$ under $\sigma$, is a
\emph{permutation} of the set $\textrm{N}.$

Let $\sigma$ and $\tau$ be two permutations of $\textrm{N}$. We
define the product $\sigma\cdot\tau$ by $(\sigma\cdot\tau)(i) =
\sigma(\tau(i))$, for $i \in \textrm{N}$, to be the composition of
the mapping $\tau$ followed by $\sigma$. These permutations taken
with $(\cdot)$ form a group denoted $S_n$ which is called the
\emph{symmetric group} of degree $n$.

Given the permutation $\sigma$ 
and for each $i \in \textrm{N}$, let us consider the sequence $i,
\sigma(i), \sigma^2(i), \dots.$ Since $\sigma$ is a bijection and
$\textrm{N}$ is finite there exist a smallest positive integer
$\ell = \ell(i)$ depending on $i$  such that $\sigma^\ell(i) = i.$
The \emph{orbit} of $i$ under $\sigma$ then consists of  the
elements $i, \sigma(i),\dots,\sigma^{\ell-1}(i)$. By a
\emph{cycle} of $\sigma$, we mean the ordered set $(i,
\sigma(i),\dots,\sigma^{\ell-1}(i))$ which sends $i$ into
$\sigma(i)$, $\sigma(i)$ into $\sigma^2(i)$,$ \dots,
\sigma^{\ell-2}(i)$ into $\sigma^{\ell-1}(i)$, and
$\sigma^{\ell-1}(i)$ into  $i$ and leaves all other elements of
\textrm{N} fixed. Such a cycle is called an $(\ell)$-cycle. We
refer to $2$-cycles as \emph{transpositions}. A pair of elements
$\{\sigma(i), \sigma(j)\}$ is  an  \emph{inversion} in a
permutation $\sigma$ if $i < j$ and $\sigma(i)>\sigma(j)$. Any
permutation can be written as a product of transpositions. The
number of transpositions in any such product is even if and only
if the number of inversions is even, and consequently, we say the
permutation is even. Similarly, a permutation is odd if it can be
written as a product of an odd number of transposition and hence
has an odd number of inversions.

\begin{lemma}\label{decomposition}
Every permutation can be uniquely expressed as a product of
disjoint cycles.
\end{lemma}
\emph{Proof:} Let $\sigma$ be a permutation. Then the cycles of
the permutation are of the form $i,
\sigma(i),\dots,\sigma^{\ell-1}(i)$. Since the cycles are disjoint
and by the multiplication of cycles, we have it that the image of
$i \in N$ under $\sigma$ is the same as the image under the
product, $\varsigma$, of all the disjoint cycles of $\sigma$.
Then, $\sigma$ and $\varsigma$ have the same effect on every
element in $N$, hence, $\sigma = \varsigma$.

Every permutation in $S_n$ has then a \emph{cycle decomposition}
that is unique up to ordering of the cycles and up to a cyclic
permutation of the elements within each cycle. Further, if $\sigma
\in S_n$ and $\sigma$ is written as the product of disjoint cycles
of length $n_1,\dots,n_k$, with $n_i\leq n_{i+1}$, we say
$(n_1,\dots,n_k)$ is the \emph{cycle type} of $\sigma$.

As a result of Lemma \ref{decomposition}, every permutation can be
written as a product of transpositions. Since  the number of
transpositions needed to represent a given permutation is either
even or odd, we  define the \emph{signature} of a permutation as
\begin{eqnarray}
\textrm{sgn}(\sigma) = \left\{%
\begin{array}{cc}
  +1 & \textrm{if $\sigma$ is even} \\
  -1  & \textrm{if $\sigma$ is odd} \\
\end{array}%
\right.
\end{eqnarray}
To each permutation,  let us associate a permutation matrix
$A_{\sigma}$ whereby \begin{eqnarray} A_{\sigma}{(j,i)} = \left\{%
\begin{array}{cc}
  1 & \textrm{if $\sigma(i)$ = $j$} \\
  0 & \textrm{otherwise} \\
\end{array}%
\right.
\end{eqnarray}
The mapping $f: S_n \mapsto \textrm{det}(A_\sigma)$ is a group homomorphism, where \begin{eqnarray}\textrm{det}(A_\sigma) = \sum_{\sigma\in S_n}{\textrm{sgn}(\sigma)\prod_{i=1}^{n}{A_{\sigma(i),i}}}\end{eqnarray}   
The kernel of this homomorphism, $\textrm{ker}f$, is  the set of
even permutations. Consequently, we have it that $\sigma \
\textrm{is even}$ if and only if $\textrm{det}(A_\sigma)$ equals
$+1$.
The kernel of the homomorphism signature defines the alternating
group. Note that the set of odd permutation can not form a
subgroup but they form a coset of the alternating group.

Let us consider the following problem. Given a pair of
$d$-dimensional quantum systems, system ${\cal{A}}$ in the state
$\qudit{\psi}$ and system ${\cal{B}}$ in the state $\qudit{\phi}$,
determine if it is possible swap the states of the corresponding
systems so that system ${\cal{A}}$ is in the state $\qudit{\phi}$
and that system ${\cal{B}}$ is in the state $\qudit{\psi}$.

\section{\underline{} Interchanging a pair of Qutrits}
Let ${\cal{H}}_{\cal{{A}}}$ and ${\cal{H}}_{\cal{{B}}}$ be two $d$-dimensional
Hilbert spaces with bases $\qudit{i}_{\cal{A}}$ and $\qudit{i}_{\cal{B}}, i \in \Z_{d}$
respectively. Let $\qudit{\psi}_A$ denote a pure state of the quantum
system ${\cal{H}}_{\cal{{A}}}$. Similarly, let $\qudit{\phi}_{\cal{{B}}}$ denote
a pure state of the quantum system ${\cal{H}}_{\cal{{B}}}$ 
and  consider an arbitrary unitary transformation $U \in {U(d}^2)$
acting on ${\cal{H}_{\cal{{A}}}} \otimes {\cal{H}_{\cal{{B}}}}$.
Let $U_{\tiny{\textrm{CNOT1}}}$ (Vatan and Williams (2004)) denote
a {\small{CNOT}} gate that has qudit $\qudit{\psi}_{\cal{{A}}}$ as
the control qudit and $\qudit{\phi}_{\cal{{B}}}$ as the target
qudit;
\begin{eqnarray}
U_{\tiny{\textrm{CNOT1}}}\qudit{m}_{{\cal{{A}}}}\otimes\qudit{n}_{{\cal{{B}}}}
= \qudit{m}_{{\cal{{A}}}}\otimes\qudit{n \oplus m}_{{\cal{{B}}}},
\qquad m,n\in \Z_d
\end{eqnarray}
 where $i\oplus j$ denote modulo $d$ addition. In gate circuitry notation, the {\small{CNOT1}} gate is given by
\begin{eqnarray}
\begin{picture}(55,25)(-25,0)
\put(-11,-3){{{\line(1,0){25}}}} \put(11,-3){{{\line(1,0){5}}}}
\put(-11,13){{{\line(1,0){25}}}} \put(11,13){{{\line(1,0){6}}}}
\put(3,-5.5){{{\line(0,1){19}}}}
\put(18,10){${\tiny{\qudit{m}_{{\cal{{A}}}}}}$}
\put(18,-5){${\tiny{\qudit{n\oplus m}_{{\cal{{B}}}}}}$}
\put(-32,-5){${\tiny{\qudit{n}_{{\cal{{B}}}}}}$}
\put(-32,10){${\tiny{\qudit{m}_{{\cal{{A}}}}}}$}
\put(3,13){\circle*{3}} \put(3,-3){\circle{5}}
\end{picture}
\end{eqnarray}
\\Similarly, let $U_{\tiny{\textrm{CNOT2}}}$   denote a {\small{CNOT}} gate that has
 qudit $\qudit{\psi}_{\cal{{A}}}$ as the target qudit and $\qudit{\phi}_{\cal{{B}}}$ as the control qudit;
\begin{eqnarray}
U_{\tiny{\textrm{CNOT2}}}\qudit{m}_{{\cal{{A}}}}\otimes\qudit{n}_{{\cal{{B}}}}
= \qudit{m\oplus n}_{{\cal{{A}}}}\otimes\qudit{n}_{{\cal{{B}}}},
\qquad m,n\in \Z_d
\end{eqnarray}
In gate circuitry notation, the {\small{CNOT2}} gate is given by
\begin{eqnarray}
\begin{picture}(55,25)(-25,0)
\put(-11,-3){{{\line(1,0){25}}}} \put(11,-3){{{\line(1,0){5}}}}
\put(-11,13){{{\line(1,0){25}}}} \put(11,13){{{\line(1,0){6}}}}
\put(3,-2){{{\line(0,1){17.5}}}} \put(18,10){$\tiny{\qudit{m\oplus
n}_{{\cal{{A}}}}}$}
\put(18,-5){${{\tiny{\qudit{n}_{{\cal{{B}}}}}}}$}
\put(-32,-5){${\tiny{\qudit{n}_{{\cal{{B}}}}}}$}
\put(-32,10){${\tiny{\qudit{m}_{{\cal{{A}}}}}}$}
\put(3,13){\circle{5}} \put(3,-3){\circle*{3}}
\end{picture}
\end{eqnarray}
\noindent We now show that a swap of two qutrits is not possible
using a composition of  {\small{CNOT}} gates alone. The point of
this argument  is to illustrate that a quantum  gate construction
which permutes the states of   three qutrit systems can not be
described by a  set of qutrit transpositions induced by the
{\small{CNOT}} gate alone. Were this otherwise then  a simple
solution to the problem of construction a generalised
{\small{SWAP}} gate for three qutrits. To argue this point, we
first note that any sequence of {\small{CNOT}} gates acting on the
qutrit states $\qudit{\psi}_{\cal{A}}$ and
$\qudit{\phi}_{\cal{B}}$ can be written as a composition of the
gates {\small{CNOT1}} and {\small{CNOT2}}. The {\small{CNOT1}} and
{\small{CNOT2}} gates can be described in the following way; the
permutation matrix corresponding to the {\small{CNOT1}} gate takes
the value 1 in row $3m+n$ and column $3m + (m\oplus n),$ $m,n =
0,1,2.$ Similarly, the matrix corresponding to the {\small{CNOT2}}
gate takes the value 1 in row $3m+n$ column $3(m\ominus n) + n$.
These unitary matrix representations for a {\small{CNOT}} gate are
given in Fig. \ref{pictures}. Furthermore, both the
{\small{CNOT1}} matrix  and {\small{CNOT2}} matrix have
determinant +1 since the permutation corresponding to each of the
respective matrices is even.

Let us now assume that there exists a gate that swaps a pair of
qutrit states and that such a gate is  composed using only the
{\small{CNOT}} gate. Such a swap gate will then  be a composition
of the gates CNOT1 and CNOT2.   Since each {\small{CNOT}} circuit
acting on a pair of qutrits is a  composition of  {\small{CNOT1}}
and {\small{CNOT2}}, it follows that any such composition will be
equivalent to some product of their respective unitary matrices.
Such a product matrix product will necessarily have  determinant
+1 as its constituent elements have determinant +1. However, the
matrix transformation representation required to effectuate the
swap of a pair of qutrits is given in Fig.~\ref{ml},
\noindent and
takes the value 1 in row $3m+n$ column $3n + m$ and has
determinant -1. Thus, no composition of the former  can yield the
latter and the result follows.

\begin{figure}
\begin{picture}(85,190)(-35,-50)
\put(-11,-3){{{\line(1,0){20}}}} \put(-11,13){{{\line(1,0){20}}}}
\put(-2,-5.5){{{\line(0,1){19}}}} \put(-2,13){\circle*{3}}
\put(-2,-3){\circle{5}}
\put(12,5){=}\put(20,10){$\left(%
\begin{array}{ccccccccc}
  1 & 0 & 0 & 0 & 0 & 0 & 0 & 0 & 0 \\
  0 & 1 & 0 & 0 & 0 & 0 & 0 & 0 & 0 \\
  0 & 0 & 1 & 0 & 0 & 0 & 0 & 0 & 0 \\
  0 & 0 & 0 & 0 & 1 & 0 & 0 & 0 & 0 \\
  0 & 0 & 0 & 0 & 0 & 1 & 0 & 0 & 0 \\
  0 & 0 & 0 & 1 & 0 & 0 & 0 & 0 & 0 \\
  0 & 0 & 0 & 0 & 0 & 0 & 0 & 0 & 1 \\
  0 & 0 & 0 & 0 & 0 & 0 & 1 & 0 & 0 \\
  0 & 0 & 0 & 0 & 0 & 0 & 0 & 1 & 0 \\
\end{array}\right)$}%
\end{picture}
\begin{picture}(85,250)(53,-170) \put(-11,-3){{{\line(1,0){20}}}}
\put(-11,13){{{\line(1,0){20}}}} \put(-2,-3.5){{{\line(0,1){19}}}}
\put(-2,-3){\circle*{3}}
\put(-2,13){\circle{5}}\put(12,5){=} \put(20,10){$\left(%
\begin{array}{ccccccccc}
  1 & 0 & 0 & 0 & 0 & 0 & 0 & 0 & 0 \\
  0 & 0 & 0 & 0 & 0 & 0 & 0 & 1 & 0 \\
  0 & 0 & 0 & 0 & 0 & 1 & 0 & 0 & 0 \\
  0 & 0 & 0 & 1 & 0 & 0 & 0 & 0 & 0 \\
  0 & 1 & 0 & 0 & 0 & 0 & 0 & 0 & 0 \\
  0 & 0 & 0 & 0 & 0 & 0 & 0 & 0 & 1 \\
  0 & 0 & 0 & 0 & 0 & 0 & 1 & 0 & 0 \\
  0 & 0 & 0 & 0 & 1 & 0 & 0 & 0 & 0 \\
  0 & 0 & 1 & 0 & 0 & 0 & 0 & 0 & 0 \\
\end{array}%
\right)$}
\end{picture}
\caption{Matrix representations of CNOT types.} \label{pictures}
\end{figure}
\section{Interchanging a pair of qudits}
Barenco \emph{et al.} (1995) showed that any unitary
transformation on a set of qubits can be decomposed into a
sequence of {\small{CNOT}} and single-qubit gates (Vidal and
Dawson (2004)). We now consider the problem of swapping a pair of
$d$-dimensional quantum states using only {\small{CNOT}} gates
such  that the system ${\cal{H}}_{\cal{{A}}}$ begins in the state
$\qudit{\psi}_{\cal{A}}$  and ends in the state
$\qudit{\phi}_{\cal{{A}}}$ while correspondingly the system
${\cal{H}}_{\cal{{B}}}$ begins in the state
$\qudit{\phi}_{\cal{{B}}}$
and ends in the state $\qudit{\psi}_{\cal{B}}$.   
 Our argument will be that a transposition of qudit states  induces some unitary matrix $U(d^2)$
 over 
${\cal{H}}_{\cal{{A}}}\otimes{\cal{H}}_{\cal{{B}}}$ whose circuit
architecture  can not be completely determined by using only
{\small{CNOT}} gates.

Recall the particular problem concerning the  swap of a pair of
qutrit systems. We have shown how the unitary matrices
$U_{\textrm{\tiny{CNOT1}}}$ and $U_{\textrm{\tiny{CNOT2}}}$ both
have  determinant +1. We also showed that  this is in contrast to
matrix $U_{\textrm{\tiny{SWAP}}}$ which describes the swapping of
states  of a pair of quantum systems where such a matrix has
determinant -1. Consequently, no composition of {\small{CNOT}}
gates alone can induce the matrix that  determines the action of
the SWAP gate. Another way to look at this is the following. The
permutations \begin{figure}
\begin{picture}(95,120)(-10,0)
\put(40,50){$\left(%
\begin{array}{ccccccccc}
  1 & 0 & 0 & 0 & 0 & 0 & 0 & 0 & 0 \\
  0 & 0 & 0 & 1 & 0 & 0 & 0 & 0 & 0 \\
  0 & 0 & 0 & 0 & 0 & 0 & 1 & 0 & 0 \\
  0 & 1 & 0 & 0 & 0 & 0 & 0 & 0 & 0 \\
  0 & 0 & 0 & 0 & 1 & 0 & 0 & 0 & 0 \\
  0 & 0 & 0 & 0 & 0 & 0 & 0 & 1 & 0 \\
  0 & 0 & 1 & 0 & 0 & 0 & 0 & 0 & 0 \\
  0 & 0 & 0 & 0 & 0 & 1 & 0 & 0 & 0 \\
  0 & 0 & 0 & 0 & 0 & 0 & 0 & 0 & 1 \\
\end{array}%
\right)$}\end{picture} \caption{$U_{\small{SWAP}}$}\label{ml}
\end{figure}  \begin{eqnarray}\sigma_{\textrm{\tiny{CNOT1}}} =
\left(%
\begin{array}{ccccccccc}
  0 & 1 & 2 & 3 & 4 & 5 & 6 & 7 & 8 \\
  0 & 1 & 2 & 4 & 5 & 3 & 8 & 6 & 7 \\
\end{array}%
\right)\nonumber\\
\sigma_{\textrm{\tiny{SWAP}}} = \left(%
\begin{array}{ccccccccc}
  0 & 1 & 2 & 3 & 4 & 5 & 6 & 7 & 8 \\
  0 & 5 & 6 & 7 & 4 & 1 & 2 & 3 & 8 \\
\end{array}%
\right)\end{eqnarray}that  correspond to the unitary matrices
$U_{\textrm{\tiny{CNOT1}}}$ and $U_{\textrm{\tiny{SWAP}}}$ have
corresponding  cycle types $(1,1,1,3,3)$ and $(1,1,1,2,2,2)$.
Hence, a {\small{CNOT}} gate fixes three basis states and permutes
the remaining states in two cycles of length 3. Each such cycle
may be written as a product of two transpositions.  Whence, the
signature of the {\small{CNOT}} permutation is +1. On the other
hand, a {\small{SWAP}} gate that swaps the states of a pairs of
qutrits contains three fixed elements and a set of three
transpositions and therefore the signature of the {\small{SWAP}}
permutation is -1 and it follows that no composition of
{\small{CNOT}} gates can lead to an execution of a swap of a pair
of qutrit systems.

More generally, a {\small{CNOT}} gate acting on a pair of
$d$-dimensional quantum systems corresponds to a permutation of
the $d^2$ basis states. We consider the case when $d = p$ is a
prime. For prime dimensions $d=p$ and taking the case of
{\small{CNOT1}}, we have it that the basis states
$\qudit{m}_{\cal{A}}\otimes \qudit{n}_{\cal{B}}$ of the system
${\cal{H}}_{\cal{AB}}$ are mapped mapped to
$\qudit{m}_{\cal{A}}\otimes \qudit{n\oplus m}_{\cal{B}}$. The
permutation associated with the  {\small{CNOT1}} mapping fixes $d$
basis states and has $(d-1)$ cycles of length $d$, each
of which may be written as a product of $d-1$ transpositions. 
{\small{CNOT1}} yields  a permutation that  can then be composed
of $(d-1)^2$ transpositions of qudit basis states. Similarly, the
{\small{CNOT2}} gate acting on a pair of qudit basis states maps
$\qudit{m}_{\cal{A}}\otimes \qudit{n}_{\cal{B}}$ of
${\cal{H}}_{\cal{AB}}$
to $\qudit{m\oplus n}_{\cal{A}}\otimes \qudit{n}_{\cal{B}}$. 
There are  $d$ fixed  basis elements under the {\small{CNOT2}}
mapping and ($d$-1) cycles, each a product of $d$-1
transpositions. 
Therefore, the  signature of the {\small{CNOT}} permutation is
$-1$ for dimension  $d=2$ and $+1$ for odd prime dimensions. Now
suppose a {\small{CNOT}} gate is acting on a pair of qudits within
system ${\cal{H}}_{d^{d}}$.  Further suppose that  such an action
is described by $U_{\textrm{\tiny{CNOT}}}\otimes I_{d^{d-2}}$.
This matrix representation induces a permutation of $d^{(d-2)}$
copies of the $d^2$ basis elements targeted by the {\small{CNOT}}
gate and it follows that the signature of corresponding
permutation is $-1$ only for dimension $d=2$.

Let us consider a SWAP gate that swaps that states of a pair of
qudits. Such a gate  corresponds to a permutation of the $d^2$
basis states of system ${\cal{H}}_{d^2}$ which  maps basis states
$\qudit{m}_{\cal{A}}\otimes \qudit{n}_{\cal{B}}$ to basis states
$\qudit{n}_{\cal{A}}\otimes \qudit{m}_{\cal{B}}$. Under this
mapping there are $d$ fixed basis elements and $d(d-1)/2$
transpositions which  describe the interchanging of all remaining
basis states.  Thus, the signature of the permutation
corresponding to  the {\small{SWAP}} gate of a pair of qudits is
$-1$ for dimensions $d \equiv 2$ or $3$ (mod $4$) and $+1$ for
dimensions $d \equiv 0$ or $1$ (mod $4$). Thus when $d \equiv 3$
(mod 4) the SWAP cannot be realised the {\small{CNOT}} gates
alone. Further consider a cycle of $d$ quantum states that maps
basis states
${\qudit{u}}_{\cal{I}}\otimes{\qudit{v}}_{\cal{J}}\otimes{\qudit{w}}_{\cal{K}}\dots\otimes{\qudit{z}}_{\cal{M}}$
to the  basis states
${\qudit{z}}_{\cal{I}}\otimes{\qudit{u}}_{\cal{J}}\otimes{\qudit{v}}_{\cal{K}}\dots\otimes{\qudit{y}}_{\cal{M}}$.
As above the cycle structure of this permutation depends on the
factorisation of the dimension of the quantum system. Thus, for
prime dimensions, the permutation corresponding to a cycle of  $d$
qudit states  contains  $d$ fixed states and $(d^d - d)/d$ cycles
of length $d$. Consequently, there are $(d^{(d-1)}-1){(d-1)}$
transpositions association with the cycle of $d$ qudit systems.
Over even dimension $d$, the permutation signature of such is $-1$
and $+1$ for odd dimension $d$.

The task of interchanging  a pair of qudit states has been argued in terms of the signature of a permutation. 
Based on this argument, we have shown that  a {\small{CNOT}} gate
acting on a pair of qudits  corresponds to a permutation whose
signature is +1, for odd prime dimensions. A   {\small{SWAP}} of
pairs of qudit systems yields a permutation whose signature is
$-1$ for dimensions $d \equiv 2$ or $3$ (mod $4$) and $+1$ for
dimensions $d \equiv 0$ or $1$ (mod $4$). By this argument alone,
circuit architectures completely described by instances of the
{\small{CNOT}} gate can not be used to implement a {\small{SWAP}}
of a pair of qudits for dimensions $ d \equiv 3$ (mod $4$).

\section{Conclusion}

We have shown that quantum circuit architectures completely
described by instances of the {\small{CNOT}} gate can not
implement a transposition of a pair of qudits for dimension $d
\equiv \ 3 \ (\textrm{mod}\ 4)$. This is of interest as
constructing  a  {\small{SWAP}} gates  for qutrits can not be
implemented through a sequence of transpositions of qutrits if
only {\small{CNOT}} gates are used.  We ask the question can  a
generalised {\small{{\small{SWAP}}}} gate for higher dimensional
quantum systems  can be  constructed entirely from instances of
the  {\small{{\small{{\small{{\small{CNOT}}}}}}}} gate.

\end{document}